\documentstyle[pre,aps,epsf,multicol]{revtex}
\begin{document}
\draft
\title{Weighted-density approximation for general nonuniform 
	fluid mixtures}
\author{Ruslan L. Davidchack and Brian. B. Laird\cite{brian}}
\address{Department of Chemistry and Kansas Institute for 
	 Theoretical and Computational Science,\\
         University of Kansas, Lawrence, Kansas 66045, USA}
\date{December 7, 1998}
\maketitle
\begin{abstract}
In order to construct a general density-functional theory for 
nonuniform fluid mixtures, we propose an extension to multicomponent 
systems of the weighted-density approximation (WDA) of Curtin and 
Ashcroft [Phys. Rev. A {\bf 32}, 2909 (1985)]. This extension corrects
a deficiency in a similar extension proposed earlier by Denton and 
Ashcroft [Phys. Rev. A {\bf 42}, 7312 (1990)], in that that functional 
cannot be applied to the multi-component nonuniform fluid systems with 
spatially varying composition, such as solid-fluid interfaces.
As a test of the accuracy of our new functional, we apply it to the
calculation of the freezing phase diagram of a binary hard-sphere 
fluid, and compare the results to simulation and the Denton-Ashcroft 
extension.
\end{abstract}

\ifpreprintsty
\else
\begin{multicols}{2}
\fi


Density-functional theory has been applied with success to 
the calculation of the equilibrium properties of a variety of 
inhomogeneous fluid systems\cite{Evans92}. One of the most successful 
of these theories has been the Weighted-Density Approximation (WDA) 
of Curtin and Ashcroft\cite{Curtin85}.  The WDA has been applied with 
good results to problems of solid-fluid phase 
equilibria\cite{Curtin85,Curtin86}, solid-liquid 
interfaces\cite{Curtin87,Ohnesorge95}, and nonuniform 
fluids\cite{Kroll90}.  
For systems with well-defined bulk densities that are uniform 
throughout (such as in the case of freezing calculations), the
Modified WDA (MWDA) of Denton and 
Ashcroft\cite{Denton89,Laird90a,Laird92} may be used, yielding 
essentially identical results as WDA with much less computation.
By construction, however, the MWDA cannot be used to study systems 
in which the bulk densities vary over length scales larger than 
a few atomic diameters (for example, interfaces or nucleation), so 
the full WDA must be utilized in such cases.  (It is, however, often 
possible to exploit the system symmetry to allow for some 
simplification of the WDA scheme, e.g. the Planar WDA of Marr and 
Gast\cite{Marr93}.)

Denton and Ashcroft have proposed an extension of the WDA and MWDA
for multi-component systems\cite{Denton90}.  Using the extended MWDA
they calculated portions of the phase diagram for the binary
hard-sphere system and obtained excellent agreement with the 
simulation results of Kranendonk and Frenkel\cite{Kranendonk89}.  
However, the WDA that they proposed assumes that the mole fractions of 
the various components of the mixture are global properties of the 
system, making their approach inapplicable to systems, such as 
interfaces,  in which spatially extended variations of composition may
exist.  In this work, we propose a different extension of WDA for 
multi-component systems that correctly preserves the local nature 
of the WDA and thus can be applied to any inhomogeneous fluid system. 
In order to verify the validity of our approximation, we also
formulate the extension of MWDA consistent with our extension of WDA 
and compare our results for freezing of binary hard-sphere mixture
with those of Denton and Ashcroft\cite{Denton90}.

We formulate our extension of WDA for the case of binary 
mixture - the extension to three or more components is
straightforward.
The state of a binary mixture is  specified by the single-particle
densities of each species $\rho_1({\bf r})$ and
$\rho_2({\bf r})$, or, alternatively, by the total density 
$\rho({\bf r}) \equiv \rho_1({\bf r}) + \rho_2({\bf r})$ and the 
mole fraction $x({\bf r})$, which we take to be the concentration
of component 2 ( $x \equiv x_2 = \rho_2/\rho$ ).  
A fundamental result of the density-functional theory is that
the Helmholtz free energy ${\cal F}[\rho_1,\rho_2]$ is a unique 
functional of the densities and is minimized (at constant average 
density) by the correct equilibrium densities. 

The free energy can be separated into ideal and excess terms
as follows
\begin{equation}
  {\cal F}[\rho_1,\rho_2] = {\cal F}_{\mathrm id}[\rho_1,\rho_2] + 
  {\cal F}_{\mathrm ex}[\rho_1,\rho_2]\:,
						\label{eq:F2}
\end{equation}
where the ideal part is simply the sum of the ideal free energies
of the individual components
\begin{equation}
  {\cal F}_{\mathrm id}[\rho_1,\rho_2] = \beta^{-1} \sum_{i = 1}^2 
  \int\!{\mathrm d}{\bf r}\,\rho_i({\bf r})
  [\ln \Lambda_i^3 \rho_i({\bf r}) - 1]\:,
						\label{eq:Fid2}
\end{equation}
with $\Lambda_i$ being the thermal wavelength of component $i$.  
An analytic expression for the excess free energy is not available 
and, in practice, this term must be approximated.  In general, a 
density-functional theory is a procedure for the specification
of this excess free energy functional.  One natural starting point for
the development of such theories is the hierarchy of $n$-particle 
direct correlation functions $c^{(n)}_{i\cdots j}$ defined as 
functional derivatives of ${\cal F}_{\mathrm ex}$ with respect to the 
densities
\begin{eqnarray}
  c^{(n)}_{i\cdots j}({\bf r}_1,\ldots,{\bf r}_n;\rho_1,\rho_2) &&
  \nonumber \\ && \hspace*{-2.5cm} = 
  - \beta\,\frac{\delta^n {\cal F}_{\mathrm ex}[\rho_1,\rho_2]}
  {\delta\rho_i({\bf r}_1)\cdots\delta\rho_j({\bf r}_n)}\:,\quad
  i,\ldots,j = 1,2\:.
\label{eq:hier12}
\end{eqnarray}
Most theories to date are constructed so that Eq.~(\ref{eq:hier12}) 
is exact in the homogeneous fluid limit - at least for $n=2$. Of such 
theories, the WDA and MWDA have been perhaps the most successful for 
single-component systems.  In what follows we discuss extensions of 
these theories to multi-component systems - the original single 
component theories can be recovered in the limit that
the mole fraction of one species becomes unity.

Denton and Ashcroft generalize the WDA to binary mixtures
by writing the WDA excess free energy in the form\cite{Denton90}
\begin{eqnarray}
  {\cal F}_{\mathrm ex}[\rho_1,\rho_2]&=&\int\!{\mathrm d}{\bf r}\,
  \rho_1({\bf r}) \psi(\bar{\rho}^{(1)}({\bf r}),x) \nonumber \\
  &&+~\int\!{\mathrm d}{\bf r}\,
  \rho_2({\bf r}) \psi(\bar{\rho}^{(2)}({\bf r}),x)\:,
						\label{eq:wd2}
\end{eqnarray}
where $\psi(\rho,x)$ is the excess free energy per particle
of the corresponding {\em uniform} mixture, and $x$ represents 
the {\em global} average concentration of the nonuniform mixture.
The two different {\em total} weighted densities
$\bar{\rho}^{(1)}({\bf r})$ and $\bar{\rho}^{(2)}({\bf r})$
are defined as weighted averages of the physical densities with
respect to weight functions $w_{ij}$ according to
\begin{eqnarray}
  \bar{\rho}^{(i)}({\bf r})&\equiv&\sum_{j = 1}^2 
  \int\!{\mathrm d}{\bf r}'
  \,\rho_j({\bf r}') w_{ij}(|{\bf r} - {\bf r}'|;
  \bar{\rho}^{(i)}({\bf r}),x)\:, \nonumber \\
  && \hspace*{4cm} i = 1,2\:.
						\label{eq:rhobwda2}
\end{eqnarray}
The weight functions must be normalized
\begin{equation}
  \int\!{\mathrm d}{\bf r}'\,w_{ij}(|{\bf r}-{\bf r}'|;\rho) 
  = 1\:,\qquad i,j = 1,2\;,
						\label{eq:wnorm2}
\end{equation}
and such that the approximate ${\cal F}_{\mathrm ex}$ exactly
satisfies Eq.~(\ref{eq:hier12}) for $n = 2$ in the uniform fluid
limit
\begin{eqnarray}
  c_{ij}(|{\bf r}_1-{\bf r}_2|;\rho_1,\rho_2) && \nonumber \\ 
  && \hspace*{-2cm} = -\beta \lim_{\rho({\bf r}) \to \rho}
  \left[\frac{\delta^2 {\cal F}_{\mathrm ex}[\rho_1,\rho_2]}
  {\delta \rho_i({\bf r}_2)\,\delta \rho_j({\bf r}_1)}\right],
  \quad i,j = 1,2\;.
						\label{eq:c2ijlim}
\end{eqnarray}
In order to compute the weight functions, one now has to solve
the system of three nonlinear differential equation\cite{Denton90}.
 
The MWDA is generalized by Denton and Ashcroft in the same spirit with
\begin{equation}
  {\cal F}_{\mathrm ex}[\rho_1,\rho_2] = 
  N_1\,\psi(\bar{\rho}^{(1)},x) + 
  N_2\,\psi(\bar{\rho}^{(2)},x)\:,
						\label{eq:fmwda2}
\end{equation}
where the weighted densities $\bar{\rho}^{(1)}$ and $\bar{\rho}^{(2)}$
are position-independent and defined as
\begin{eqnarray}
  \bar{\rho}^{(i)}&\equiv&\frac{1}{N_i} \sum_{j = 1}^2 
  \int\!{\mathrm d}{\bf r}\,{\mathrm d}{\bf r}'\,
  \rho_i({\bf r}) \rho_j({\bf r}') w_{ij}(|{\bf r} - {\bf r}'|;
  \bar{\rho}^{(i)},x)\:, \nonumber \\
  && \hspace*{5cm} i = 1,2\:.
						\label{eq:rhobmwda2}
\end{eqnarray}
As in the single-component case, the MWDA is much less
computationally demanding, so Denton and Ashcroft have used this 
approximation to compute freezing conditions of the binary 
hard-sphere mixtures with different diameter ratios.  Their results
for $\alpha > 0.85$ closely follow those obtained in the simulations
by Kranendonk and Frenkel \cite{Kranendonk89}.

Nevertheless, the binary mixture WDA of Denton and Ashcroft defined
by Eqs.~(\ref{eq:wd2}) and (\ref{eq:rhobwda2}) cannot be used to
study any binary system with extended spatial variations of average
composition. The problems are best illustrated by the example of
a planar crystal-fluid interface. The excess free energy 
${\cal F}_{\mathrm ex}$ in Eq.~(\ref{eq:wd2}) and, therefore, the 
weight functions $w_{ij}$ in Eq.~(\ref{eq:rhobwda2}) are defined to 
depend explicitly on the global average concentration $x$.  Obviously 
this quantity cannot be uniquely defined for the interfacial system, 
where it has to be equal to the crystal coexistence values 
$x^{\mathrm c}$ on the one side of the interface and to the fluid 
value $x^{\mathrm f}$ on the other side.  Matematically, the WDA 
equations for  $w_{ij}(k;\rho,x)$ derived by Denton and 
Ashcroft\cite{Denton90} contain terms proportional to $\delta_{k,0}$,
so that the computed weight functions are discontinuous in
Fourier space at $k = 0$.  This means that one cannot recover the 
bulk crystal or fluid properties in the regions remote from the 
interface (in the limit $z \to \pm \infty$, where $z$ is the axes 
perpendicular to the interfacial plane).  

In order to circumvent these difficulties, we propose an extension 
of the WDA to binary mixtures in which {\em weighted} concentrations 
$\bar{x}^{(1)}({\bf r})$ and $\bar{x}^{(2)}({\bf r})$ are introduced 
in addition to the weighted densities.  The weighted concentrations 
replace the average concentration $x$ in Eq.~(\ref{eq:wd2}) and are 
defined according to
\begin{eqnarray}
  \bar{x}^{(i)}({\bf r})&\equiv&\frac{1}{\bar{\rho}^{(i)}({\bf r})}
  \int\!{\mathrm d}{\bf r}' \,\rho_2({\bf r}') \nonumber \\ 
  &&\times w_{i2}(|{\bf r} - {\bf r}'|;\bar{\rho}^{(i)}({\bf r}),
  \bar{x}^{(i)}({\bf r}))\:,\quad i = 1,2\;.
						\label{eq:xbwda2}
\end{eqnarray}
It is more convenient to write this extension scheme in terms of
the weighted densities of individual species 
$\bar{\rho}^{(i)}_j({\bf r})$ defined as
\begin{eqnarray}
  \bar{\rho}^{(i)}_j({\bf r})&\equiv&\int\!{\mathrm d}{\bf r}'
  \,\rho_j({\bf r}') w_{ij}(|{\bf r} - {\bf r}'|;
  \bar{\rho}^{(i)}_1({\bf r}),
  \bar{\rho}^{(i)}_2({\bf r}))\;, \nonumber \\
  && \hspace*{4cm} i,j = 1,2\;,
						\label{eq:rhobijwda2}
\end{eqnarray}
which are input into the excess free energy expression
\begin{eqnarray}
  {\cal F}_{\mathrm ex}[\rho_1,\rho_2]&=&\int\!{\mathrm d}{\bf r}\,
  \rho_1({\bf r}) \psi(\bar{\rho}^{(1)}_1({\bf r}),
  \bar{\rho}^{(1)}_2({\bf r})) \nonumber \\
  &&+~\int\!{\mathrm d}{\bf r}\,
  \rho_2({\bf r}) \psi(\bar{\rho}^{(2)}_1({\bf r}),
  \bar{\rho}^{(2)}_2({\bf r}))\:,
						\label{eq:wdij2}
\end{eqnarray}
where $\psi(\rho_1,\rho_2)$ is the same excess free energy per 
particle as in Eq.~(\ref{eq:wd2}), but expressed in variables 
$\rho_1$ and $\rho_2$.

Substitution of ${\cal F}_{\mathrm ex}$ from Eq.~(\ref{eq:wdij2})
into Eq.~(\ref{eq:c2ijlim}) leads explicitly to the following
set of equations for the three weight functions in Fourier space
\begin{eqnarray}
  -\beta^{-1} c_{11}(k)&=&2 \frac{\partial \psi}
  {\partial \rho_1}w_{11}(k) + 
  \frac{\partial^2 \psi}{\partial \rho^2_1}\Big[\rho_1 w^2_{11}(k) +
  \rho_2 w^2_{12}(k)\Big] \nonumber \\ &&\hspace*{-1.2cm} 
  +~2\frac{\partial \psi}{\partial \rho_1}\Big[\rho_1
  w_{11}(k)\frac{\partial w_{11}(k)}{\partial \rho_1} + \rho_2
  w_{12}(k)\frac{\partial w_{12}(k)}{\partial \rho_1}\Big]\:,
  \nonumber \\ 
		&&				\label{eq:w11k}
\end{eqnarray}
\begin{eqnarray}
  -\beta^{-1} c_{12}(k)&=&\left( \frac{\partial \psi}
  {\partial \rho_1} + \frac{\partial \psi}
  {\partial \rho_2}\right) w_{12}(k) \nonumber \\
  &&\hspace*{-1.2cm} +~\frac{\partial^2 \psi}
  {\partial \rho_1\,\partial \rho_2}\Big[\rho_1 w_{11}(k) + 
  \rho_2 w_{22}(k)\Big]w_{12}(k) \nonumber \\
  &&\hspace*{-1.2cm} +~\frac{\partial \psi}{\partial \rho_1}\Big[\rho_1
  w_{12}(k)\frac{\partial w_{11}(k)}{\partial \rho_2} + \rho_2
  w_{22}(k)\frac{\partial w_{12}(k)}{\partial \rho_2}\Big]
  \nonumber \\
  &&\hspace*{-1.2cm} +~\frac{\partial \psi}{\partial \rho_2}\Big[\rho_1
  w_{11}(k)\frac{\partial w_{12}(k)}{\partial \rho_1} + \rho_2
  w_{12}(k)\frac{\partial w_{22}(k)}{\partial \rho_1}\Big]\:,
  \nonumber \\
  		&&				\label{eq:w12k}
\end{eqnarray}
\begin{figure}
\epsfysize=8cm \epsfbox{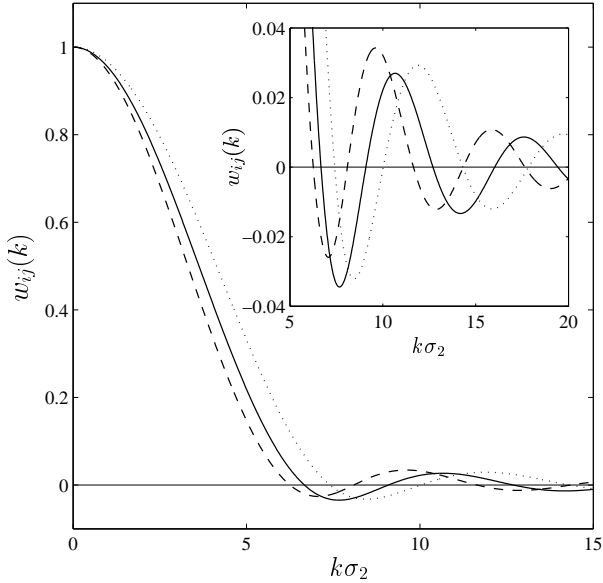}
\caption[WDA weight functions $w_{ij}(k)$ for a binary hard-sphere
	mixture.]
	{\narrowtext 
	WDA weight functions $w_{ij}(k)$ for a binary hard-sphere
	mixture.   The system parameters are: 
	$\alpha = 0.8\,$, $x = 0.7\,$, and 
        $\rho = 1.1\,\sigma_2^{-3}$. 
	The dotted, solid, and dashed curves correspond to 
	$w_{11}\,$, $w_{12}\,$, and $w_{22}\,$, respectively.}
						\label{fig:b9}
\end{figure}   
\begin{eqnarray}
  -\beta^{-1} c_{22}(k)&=&2 \frac{\partial \psi}
  {\partial \rho_2}w_{22}(k) + 
  \frac{\partial^2 \psi}{\partial \rho^2_2}\Big[\rho_1 w^2_{12}(k) +
  \rho_2 w^2_{22}(k)\Big]\nonumber \\ &&\hspace*{-1.2cm} 
  +~2\frac{\partial \psi}{\partial \rho_2}\Big[\rho_1
  w_{12}(k)\frac{\partial w_{12}(k)}{\partial \rho_2} + \rho_2
  w_{22}(k)\frac{\partial w_{22}(k)}{\partial \rho_2}\Big]\:.
  \nonumber \\
		&&				\label{eq:w22k}
\end{eqnarray}
The normalization of the weight functions ensures that at $k = 0$ 
Eqs.~(\ref{eq:w11k})-(\ref{eq:w22k}) correctly reduce to the
compressibility rules for a binary mixture\cite{Ashcroft67}
\begin{eqnarray}
  -\beta^{-1} c_{ij}(k = 0;\rho_1,\rho_2)&=& 
  \frac{\partial \psi}{\partial \rho_i} +
  \frac{\partial \psi}{\partial \rho_j} + \rho\frac{\partial^2 \psi}
  {\partial \rho_i\,\partial \rho_j}\:,\nonumber \\
  &&\hspace*{2cm} i,j = 1,2\;.
						\label{eq:crule2}
\end{eqnarray}
Unlike the original generalization of Denton and 
Ashcroft\cite{Denton90}, the above equations do not contain terms 
proportional to $\delta_{k,0}$, and are thus continuous 
functions of $k$.  

In order to calculate the weight functions from 
Eqs.~(\ref{eq:w11k})-(\ref{eq:w22k}), we need to specify $c_{ij}(k)$, 
the Fourier transforms of the uniform direct correlation functions, 
and $\psi(\rho_1,\rho_2)$, the Helmholtz free energy per particle.  
As an example, we have calculated the weight functions for
the binary hard-sphere mixture with diameter ratio 
$\alpha \equiv \sigma_1/\sigma_2 = 0.8$ and use as input 
the analytical expressions for a uniform binary mixture 
within the Percus-Yevick approximation\cite{Liebowitz64,Ashcroft67}.  
 
Eqs.~(\ref{eq:w11k})-(\ref{eq:w22k}) is a system of nonlinear
differential equations.  We use an iterative scheme, which
has satisfactory convergence for all values of $k$ and density.
We start by assuming that $w_{12}$ and all derivatives of $w_{ij}$ are
zero and solve the quadratic equations for $w_{11}$ and $w_{22}$.  
Then we estimate the derivatives of $w_{11}$ and $w_{22}$ using finite
differences and solve equation for $w_{12}$, which is linear 
in $w_{12}$.  Finally, we estimate the derivative of $w_{12}$.
After 5--10 iterations the solution converges to within five
significant digits.  Fig.~\ref{fig:b9} shows $w_{ij}(k)$ computed
for the binary hard-
\begin{figure}
\hspace*{-1cm} 
\epsfxsize=9cm \epsfbox{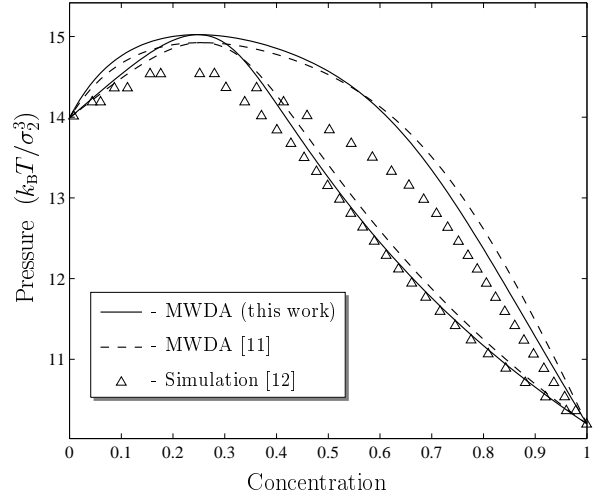}
\vspace*{.2cm}
\caption[Pressure-concentration phase diagrams for the 
	binary hard-sphere system with diameter ratio $\alpha = 0.9$.]
	{\narrowtext
	Comparison of binary hard-sphere crystal-melt phase
	diagrams computed with the original MWDA by
	Denton and Ashcroft {\protect \cite{Denton90}} (dashed lines) 
        and our new extension of the MWDA (solid lines).  Also 
	shown are the {\em rescaled} simulation data
	from Ref. {\protect \cite{Kranendonk89}} (see text for
	explanation).}
						\label{fig:b10}
\end{figure}   
\noindent sphere fluid with
diameter ratio $\alpha = 0.8\,$ concentration $x = 0.7\,$, and 
total density $\rho = 1.1\,\sigma_2^{-3}$.  All three functions
smoothly approach unity when $k \to 0$, thus satisfying the
normalization conditions and allowing application of the
WDA to binary systems with spatially varying characteristics.  

In order to compare our extension of the WDA to mixtures with
that of Denton and Ashcroft \cite{Denton90}, we repeat the
freezing calculations for binary hard-sphere mixture using
MWDA consistent with the extension of the WDA presented earlier.
The excess free-energy functional is written as
\begin{equation}
  {\cal F}_{\mathrm ex}[\rho_1,\rho_2] = 
  N_1\,\psi(\bar{\rho}^{(1)}_1,\bar{\rho}^{(1)}_2) + 
  N_2\,\psi(\bar{\rho}^{(2)}_1,\bar{\rho}^{(2)}_2)\:,
						\label{eq:fmwda3}
\end{equation}
where the weighted densities $\bar{\rho}^{(i)}_j$ are defined as
\begin{eqnarray}
  \bar{\rho}^{(i)}_j&\equiv&\frac{1}{N_i} 
  \int\!{\mathrm d}{\bf r}\,{\mathrm d}{\bf r}'\,
  \rho_i({\bf r}) \rho_j({\bf r}') w_{ij}(|{\bf r} - {\bf r}'|;
  \bar{\rho}^{(i)}_1,\bar{\rho}^{(i)}_2)\:, \nonumber \\
  &&\hspace{5cm} i,j = 1,2\;.
						\label{eq:rhobmwda3}
\end{eqnarray}
This extension of MWDA yields the following equations for the
weight function $w_{ij}(k)$ in Fourier representation
\begin{eqnarray}
  -\beta^{-1} c_{ij}(k)&=&\left( \frac{\partial \psi}
  {\partial \rho_i} + \frac{\partial \psi}
  {\partial \rho_j}\right) w_{ij}(k) + \delta_{k,0} 
  \rho \frac{\partial^2 \psi}{\partial \rho_i \partial \rho_j}\;,
  \nonumber \\ &&\hspace*{4cm} i,j = 1,2\;,
						\label{eq:wijk}
\end{eqnarray}
where $\rho = \rho_1 + \rho_2$ is the total density.  Agreement with
the compressibility rules given by Eq.~\ref{eq:crule2} is obvious.

Comparison between the freezing parameters obtained within our 
extension of the MWDA and that of Denton and Ashcroft is 
shown in Figs.~\ref{fig:b10}-\ref{fig:b12}.  Fig.~\ref{fig:b10} 
presents the pressure-concentration phase diagrams together with 
the simulation results of Kranendonk and Frenkel\cite{Kranendonk89},
which are scaled down by a factor of 0.871 in order to match 
\begin{figure}
\hspace*{.3cm} 
\epsfysize=8cm \epsfbox{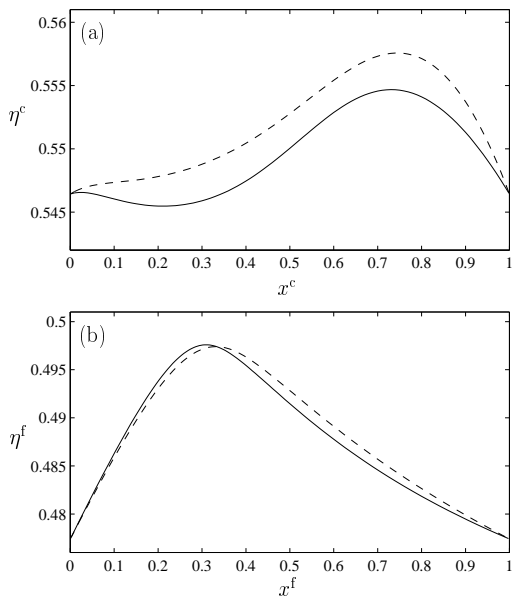}
\vspace*{.2cm}
\caption[Crystal and fluid packing fractions at coexistence.]
	{\narrowtext
	(a) Packing fraction in the crystal phase at coexistence.
	 Dashed lines represent the results of the original MWDA
	 and solid lines represent the results from our
	 extension of the MWDA.
	 (b) Packing fraction in the fluid phase at coexistence.}
						\label{fig:b11}
\end{figure}   
\noindent the 
transition parameters in the single-component limit ($x = 0$ and 1).
The scaling compensates for the discrepancies between theory and 
simulation in the freezing parameters of the one-component system 
and helps to emphasize the general features of the binary system phase
diagram.  Overall, the two extensions of the MWDA produce very
similar phase diagrams and reproduce fairly well qualitative behavior
of the simulation results.

Fig.~\ref{fig:b11} compares the crystal and fluid packing
fractions at coexistence.  The packing fractions in the crystal 
phase differ for the two extensions much more than the fluid
packing fractions.  

Fig.~\ref{fig:b12} shows the Lindemann parameters for the two types
of spheres.  In both cases, the smaller spheres have larger Lindemann 
parameters for all values of concentration. 
The Lindemann parameters for the two extensions of MWDA coincide 
for larger spheres at $x = 1$, and for smaller spheres at $x = 0$, 
where the single-component limit is recovered for the remaining
type of spheres.

In conclusion, we have constructed an extension of the 
Weighted-Density Approximation (WDA) of Curtin and 
Ashcroft\cite{Curtin85} to multi-component systems.  This new 
density-functional theory is more general than a previous extension
by Denton and Ashcroft\cite{Denton90} in that it
preserves the local character of the weighted densities and 
thus allows its application to the systems with extended spatial
variations in average composition, such as solid-liquid interfaces.

This research was supported by the National Science Foundation under 
the grant CHE-950281.  The authors also thank the Kansas Insitute 
for Theoretical and Com-
\begin{figure}
\epsfxsize=8cm \epsfbox{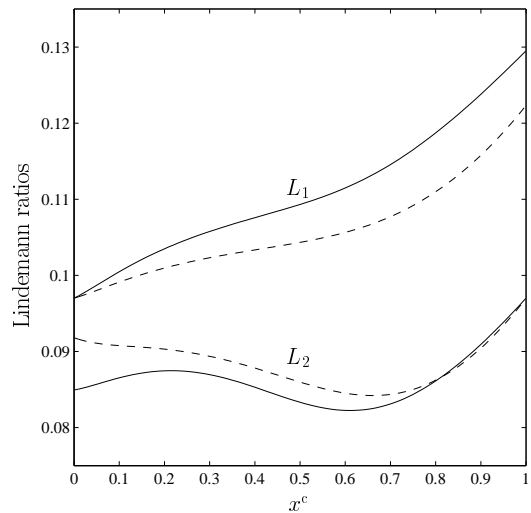}
\vspace*{.2cm}
\caption[Lindemann parameters for the two types of spheres.]
	{\narrowtext
	Lindemann parameters for the two types of spheres
	calculated according to the original MWDA (dashed lines)
	and our extension (solid lines).}
						\label{fig:b12}
\end{figure}   
\noindent putational Science (KITCS) and the Kansas 
Center for Advanced Scientific Computing (KCASC) for support
and computing resources.

\ifpreprintsty
\else
\end{multicols}
\fi


\begin{references}
\bibitem[*]{brian} Author to whom correspondence should be addressed.
\end{references}
\end{document}